\documentstyle[twoside,fleqn,epsf,espcrc2]{article}
 
\newcommand{\AmS}{{\protect\the\textfont2
  A\kern-.1667em\lower.5ex\hbox{M}\kern-.125emS}}
 
\title{ The PZ method for estimating determinant ratios, with applications}
 
\author{C. Thron, K. F. Liu, and  S. J. Dong \address{Department of Physics
and Astronomy, University of Kentucky, Lexington, KY 40506, USA}
 \thanks{This work is partially supported by DOE grant
        No. DE-Fg05-84ER40154.}}

\begin{document}
 
\begin{abstract}
We introduce a new method for estimating determinants or
determinant ratios of large matrices, which combines the techniques of
Pad\`{e} approximation with rational functions and $Z_{2}$ noise estimation
of traces of large matrices.  The method requires simultaneously solving
several matrix equations, which can be conveniently accomplished using
the MR method.  We include some preliinary results, and indicate
potential applications to non-Hermitian matrices, and Hybrid Monte Carlo
without pseudofermions.
\end{abstract}
 
\maketitle
 
\section{Introduction}
 
The current work was motivated by our search for an alternative
implementation of the Hybrid Monte Carlo (HMC) algorithm in Lattice
QCD  which does not require the use of pseudofermions  Our proposed
algorithm (which is described in more detail in \cite{Liu96} requires the
estimation of determinant ratios $ \det {\bf M}_{1}/ \det {\bf M}_{2}$
where the matrices ${\bf M}_{1}$ and ${\bf M}_{2}$ satisfy the following
conditions: ${\bf M}_{1}$ and ${\bf M}_{2}$ are huge ($N \times N$,
where $N \sim 10^{6}$); the eigenvalues of ${\bf M}_{1}$ and ${\bf M}_{2}$
have positive real  parts; $log(\det{\bf M}_{i})$ are $O(N)$, but $\det {\bf
M}_{1} /det {\bf M}_{2}$  is $O(1)$; and the eigenvalues of ${\bf M}_{2}$
are continuous perturbations of the eigenvalues of  ${\bf M}_{1}$.
 
In this paper we present a new method for estimating such determinant
ratios, which takes advantage of Pad\`{e} approximation and $Z_{2}$
noise vectors (hence the acronym 'PZ'). The potential application of
related methods to estimating the density of states. is also indicated.
 
\section{Outline of the PZ Method}
 
Our improved algorithm is based on the approximation of $\log \lambda$
with a rational function via the use of Pad\`{e} approximants. The
Pad\`{e} approximation to $ \log (z)$  of order $[J,J]$ at $z_{0}$ is a
rational function $N(z) / D(z)$ where deg$N(z)$ = deg$D(z)$ = $J$,
whose value and first $2J$  derivatives agree with $\log z$ at the
specified point $z_{0}$. When the Pad\`{e} approximation is expressed in
partial fractions, we obtain
 
\begin{equation} \label{log_M}
\log {\bf M} \approx a_{0} {\bf I} + \sum_{j=1}^{J} a_{j}({\bf M} +
b_{j})^{-1},
\end{equation}
which implies
\begin{eqnarray} \label{det_ratio}
\log \left\{ \det {\bf M}_{1}  / det {\bf M}_{2} \right\} = Tr \left[ \log {\bf M}_{1} -
\log {\bf M}_{2} \right] \nonumber \\
=\sum_{j=1}^{J} a_{j} \left\{Tr({\bf M}_{1} + b_{j})^{-1} -
                                        Tr({\bf M}_{2}+b_{j})^{-1} \right\}.
\end{eqnarray}
 
The traces of inverse matrices $( {\bf M} + b_{j} )^{-1}$ may then be
found using the $Z_{2}$ noise method \cite{Dong94}, and it follows
 
\begin{equation} \label{Tr_log}
Tr \left[ \log  {\bf M}_{1} - \log {\bf M}_{2} \right] \approx \frac{1}{L}  \sum_{j,l}
a_{j} \eta_{l}^{\dagger} (\xi_{j,l}^{(1)} - \xi_{j,l}^{(2)}),
\end{equation}
where $\left\{\eta_{l}, l=1,\ldots ,L \right\}$ are complex $Z_{2}$ noise vectors, and
the $\xi_{j,l}^{(k)}$ are vectors defined  by:
 
\begin{equation} \label{xi_def}
\xi_{j,l}^{(k)} = ( {\bf M}_{k} + b_{j} {\bf I} )^{-1} \eta_{l}.
\end{equation}
 
\section{Advantages of the PZ method}
 
The PZ method takes advantage of proven, effective numerical
approximation techniques. Pad\`{e} approximation uses rational functions,
which are very efficient when it comes to uniform approximation of
analytic functions:  for example, an $[11,11]$ Pad\`{e} expansion of
$\log z$ around  $z_{0} =.1$ is accurate to within $10^{-7}$ on the
interval $[.004,2.5]$. In our application, the Pad\`{e} approximation to
the logarithm only needs to be accurate on the region in the complex
plane where the spectra of ${\bf M}_{1}$  and ${\bf M}_{2}$ differ.  Also,
complex $Z_{2}$ random vectors have been shown to be superior to
Gaussian in computing traces of inverse matrices.
 
The PZ method is also in a position to take advantage of recent results
on
efficient solution of linear problems with multiple diagonal shifts.
The
column inversions in  \ref{xi_def} for fixed $l,k$ may be performed
using
GMRES with multiquarks  in the same computational time it takes to
solve just one--only more memory is required (an additional vector for
each j) \cite{Glassner96}.   This can lead to a considerable speedup in
the
algorithm.  Hence, a higher order Pad\`{e} expansion requires more
memory,  but the same computation time (apart from matrix conditioning
effects, see next paragraph). The multi-quark GMRES method can be
applied to non-Hermitian matrices:  so determinants of non-Hermitian
matrices may also be found directly, without recourse to the Hermitian
matrix ${\bf M}^{\dagger}{\bf M}$.
 
 The  $b_{j}$'s turn out to have positive real parts, and negligable
imaginary parts:  thus the matrices ${\bf M}+b_{j}{\bf I}$ are better-
conditioned than ${\bf M}$, and the column inversions should be faster.
 
However, this effect diminishes for higher order
Pad\`{e} approximations,
because the real parts of some terms become smaller.
 
The PZ method also holds promise of being useful in the case where
different quark flavors are present, in which case  it is necessary to
compute
 
\begin{equation}
\frac{\det \left\{ \left( {\bf M}_{1} + \mu_{1} \right)
 \left( {\bf M}_{1} + \mu_{2} \right) \right\}}
{\det  \left\{ \left( {\bf M}_{2} + \mu_{1} \right)
                  \left( {\bf M}_{2} + \mu_{2} \right) \right\}}
\end{equation}
 
Using multi-quark GMRES, this takes the same amount of computation
as a single determinant ratio.
 
\section{ Numerical Results}
 
Figure 1 shows the result of numerical experiments involving the matrix
${\bf M}_{1}$ associated with a typical (heatbath-generated) quenched
gauge configuration on a $16^3 \times 24$ lattice with $\beta=6.0$,
$\kappa=0.154$, and the matrix ${\bf M}_{2}$  generated from ${\bf
M}_{1}$ by updating 20 links in a heatbath.  The PZ estimates of
$\log | \det {\bf M}_{1} / \det {\bf M}_{2} |$ using  a $[5,5]$
Pad\`{e}
approximation with up to 50 $Z_2$ noise vectors is shown in Figure 1, and the
jackknife error is shown in Figure 2.  The PZ estimate of the real part
of
$log(\det{\bf M}_{1})$ (using (1) with an additional term $a_{0}N$ and
$\xi_{j,l}^{(2)}=0$) is shown in Figure 3, while the corresponding
jackknife errors are shown in Figure 4. The PZ estimate of the
imaginary
part of $log(\det{\bf M}_{1})$ is shown in Figure 5 (the actual value is
0).
 
These results demonstrate that the PZ method leads to controlled error
in
determinant ratios after a relatively small number of column
inversions.
 
\begin{figure}[htb]
\vspace{9pt}
\setlength\epsfxsize{70mm}
\epsfbox{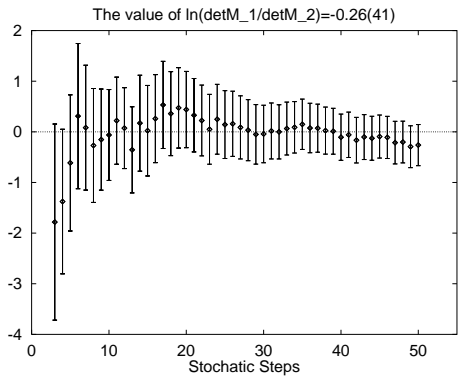}
\caption{ PZ estimates of determinant ratio}
\end{figure}
 
\begin{figure}[htb]
\vspace{9pt}
\setlength\epsfxsize{70mm}
\epsfbox{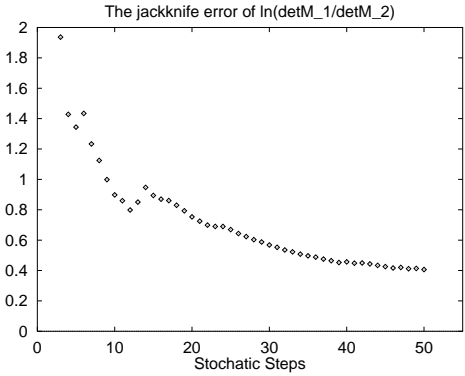}
\caption{ Jackknife error for estimates in Figure 1.}
\end{figure}
 
\section{Numerical error and computation time in the PZ method}
 
There are three sources of numerical error in the PZ method:  Pad\`{e}
approximation; $Z_{2}$ estimation; and column inversion via GMRES.
The Pad\`{e} error can be virtually eliminated by taking more terms in
the Pad\`{e} expansion, which requires more memory but not more
computation time.   A theoretical expression for the  $Z_{2}$  error is
 
\cite{Bernardson94}:
 
\begin{equation}
\mbox{Var} (Tr {\bf A} - Tr \eta^\dagger {\bf A}\eta) =
\frac{1}{L} \sum_{I \neq  j} \| {\bf A}_{ij} \| ^{2}.
\end{equation}
The column inversion error is reduced by taking more iterations, at the
cost of increased computation time.
 
\begin{figure}[ht]
\vspace{9pt}
\setlength\epsfxsize{70mm}
\epsfbox{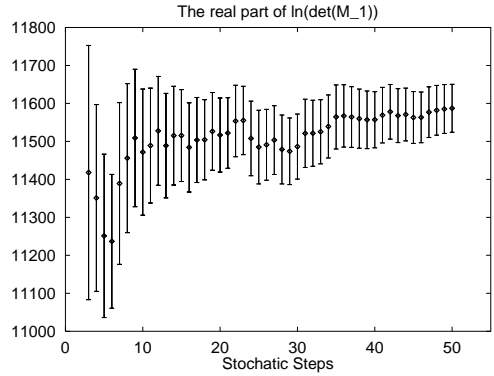}
\caption{ PZ estimate of real part of $log(\det{\bf M})$.}
\end{figure}
 
\begin{figure}[ht]
\vspace{9pt}
\setlength\epsfxsize{70mm}
\epsfbox{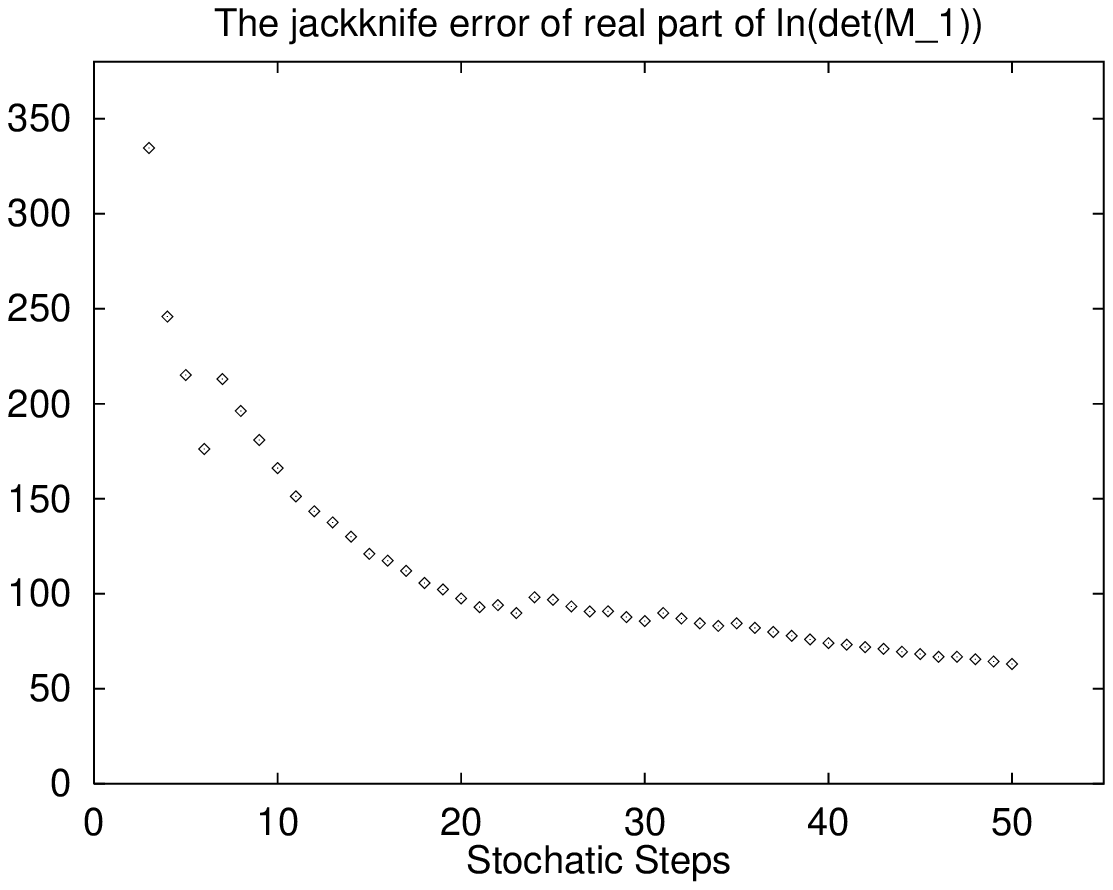}
%%\framebox[55mm]{\rule[-21mm]{0mm}{43mm}}
\caption{ Jackknife error for estimates in Figure 3.}
%%\label{fig:jackknife_error2}
\end{figure}
 
\begin{figure}[htb]
\vspace{9pt}
\setlength\epsfxsize{70mm}
\epsfbox{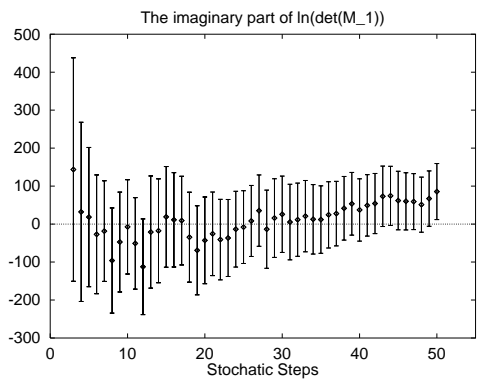}
%%\framebox[55mm]{\rule[-21mm]{0mm}{43mm}}
\caption{ PZ estimate of imaginary part of $log(\det{\bf M})$.}
%%\label{fig:det_im_part}
\end{figure}
 
The computation time depends on the number of noise vectors required
to get a good trace estimate. If GMRES is used, each noise vector used
gives rise to one column inversion.
 
 It should be mentioned that alternatives to Pad\`{e} approximation
have
been proposed, including Chebyshev polynomials \cite{Sexton95} and
Stieltjes integrals \cite{Bai96}. Numerical experiments should be
performed to determine which is more efficient.
 
\section{Estimating the density of states}
 
Some of the ideas introduced above may be used to estimate the density
of states $\rho(\lambda)$ as follows.  Any term of the form
 
\begin{eqnarray}
 Tr \{ ( {\bf M} - b_{j} )^{-1} \} = \sum_{n=1}^{N}
\frac{1}{\lambda_{n} - b_{j} } \nonumber \\ 
= \int \int {d^{2} \lambda  \frac{1}{\lambda - b_{j} }} \rho (\lambda)
\end{eqnarray}
can be used as a "probe" to sample spectral information near the
specified
pole $b_{j}$. Hence $\rho(\lambda)$ can be estimated via the following
procedure: (1)  Choose $J$ complex numbers {$b_{j}$} which are
strategically placed around the support of  $\rho (\lambda)$; (2)
Estimate $T_{j} \equiv Tr \{ ( {\bf M} - b_{j} ) ^{-1} \}$ using the
$Z_{2}$ noise method; (3) Specify a fitting form for $\rho (\lambda)$
using
$J$ parameters: $\rho (\lambda) \approx \sum_{j=1}^{J} p_{j} \phi_{j}
(\lambda)$; and (4) Find the parameters $p_j$ by solving the equations
 
\begin{equation}
T_{j} = \sum{j=1}{J} p_{j} \int \int[d^2 \lambda]
\frac{\phi_{j} (\lambda)}{\lambda - b_{j}} .
\end{equation}

\end{document}